%% file: Su_cdc.tex
\documentclass[12pt]{article}
\usepackage[margin=1in]{geometry}
\input{./sections/header}

\begin{document}
\title{\bf A Hybrid Queuing Model for Coordinated Vehicle Platooning on Mixed-Autonomy Highways: Training and Validation}

\author{Haoran Su, Zhengjie Ji, Karl. H. Johansson and Li Jin
\thanks{This work was supported in part by NYU Tandon School of Engineering, C2SMART University Transportation Center, D.D. Eisenhower Transportation Fellowship, SJTU-UM Joint Institute, J. Wu \& J. Sun Endowment Fund.}
\thanks{H. Su is with the Tandon School of Engineering, New York University, USA.
Z. Ji and L. Jin are with the UM Joint Institute, Shanghai Jiao Tong University, China. 
Z. Ji is also and K.H. Johansson is with the School of Electrical Engineering and Computer Science, KTH Royal Institute of Technology, Sweden.
L. Jin is also with the School of Electronic Information and Electrical Engineering, Shanghai Jiao Tong University, China and the Tandon School of Engineering, New York University, USA
(emails: hs1854@nyu.edu, jizhengjie@sjtu.edu.cn, kallej@kth.se, li.jin@sjtu.edu.cn).}%
}

\maketitle

\begin{abstract}
Platooning of connected and autonomous vehicles (CAVs) is an emerging technology with a strong potential for throughput improvement and fuel reduction. Adequate macroscopic models are critical for system-level efficiency and reliability of platooning.
In this paper, we consider a hybrid queuing model for a mixed-autonomy highway section and develop an easy-to-use training algorithm.
The model predicts CAV and non-CAV counts according to the traffic demand as well as key parameters of the highway section.
The training algorithm learns the highway parameters from observed data in real time.
We test the model and the algorithm in Simulation of Urban Mobility (SUMO) and show that the prediction error is around 15\% in a stationary setting and around 25\% in a non-stationary setting.
We also show that the trained model leads to a platoon headway regulation policy very close to the simulated optimum.
The proposed model and algorithm can directly support model-predictive decision making for platooning in mixed autonomy.
\end{abstract}


\input{./sections/introduction} 
\input{./sections/model} 
\input{./sections/algorithm} 
\input{./sections/validation} 
\input{./sections/conclusion} 

\bibliographystyle{IEEEtran}
\bibliography{Su_cdc}   
\end{document}

%% file: sections/header.tex
\usepackage{booktabs}
\usepackage{amsmath}
\usepackage[pdftex]{graphicx}

\usepackage{epstopdf}
\usepackage{multirow}
\usepackage{mdwlist}

\usepackage{threeparttable}

\usepackage{amsfonts}
\usepackage{amsmath}
\usepackage{hyperref}
\usepackage{float}
\usepackage{caption}
\usepackage{subcaption}
\usepackage[ruled,vlined]{algorithm2e}

%% file: sections/introduction.tex
\section{Introduction}

\begin{figure}[hbtp]
    \begin{subfigure}{\linewidth}
    \centering
    \includegraphics[width=0.75\textwidth]{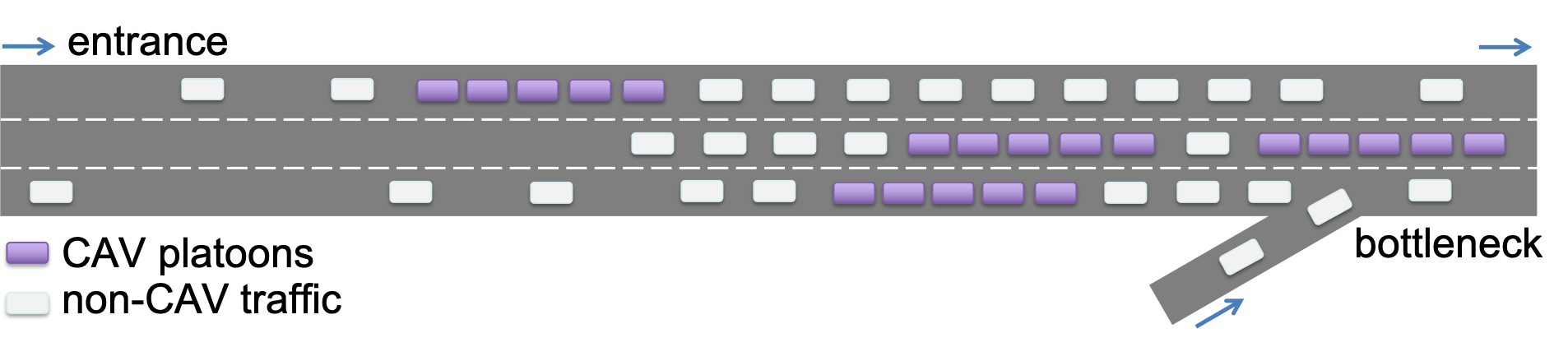}  
    \caption{Cluster of platoons may disrupt local traffic.}
    \end{subfigure}
    \begin{subfigure}{\linewidth}
    \centering
    \includegraphics[width=0.75\textwidth]{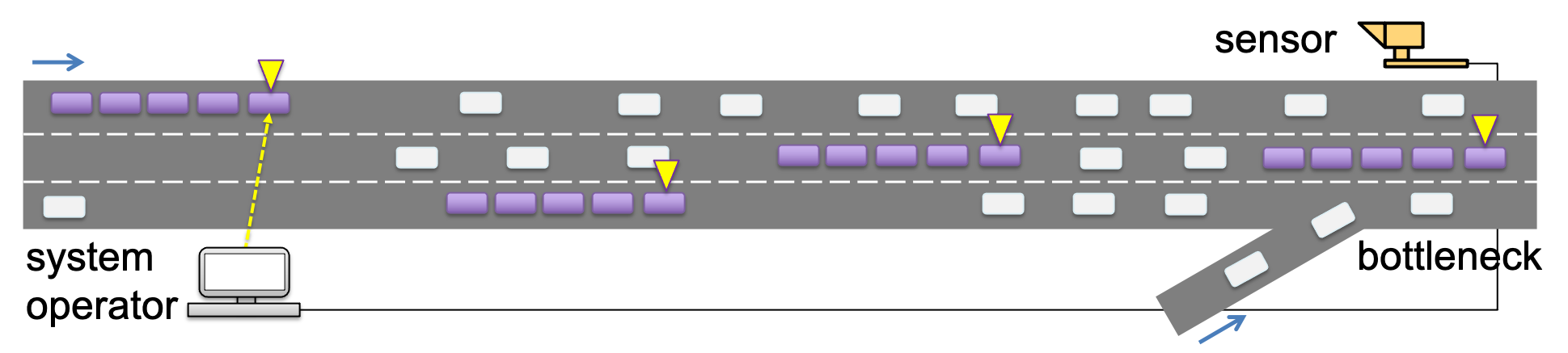}  
    \caption{Headway regulation to avoid clustering.}
    \end{subfigure}
    \caption{Analysis and design of coordinated platooning requires a macroscopic model capturing interaction between CAV platoons and non-CAV traffic.}
    \label{fig_background}
\end{figure}

Platooning of connected and autonomous vehicles (CAVs) is an emerging highway operation with a strong potential for throughput improvement and fuel savings~\cite{hor+var00,bess+16procieee,litman2017autonomous}. Platooning technology has been progressing fast recently~\cite{naus+10,ploeg14,coogan15interconnected,tsugawa16}. 
The mixed-autonomy traffic flow stability and throughput are extensively studied~\cite{yu2020stability, zhou2020modeling, wu2017emergent, talebpour2016influence}.
However, the macroscopic interaction between CAV platoons and non-CAV background traffic has not been well modeled or understood.
In particular, platoons require much larger spaces for lane-changing than non-CAVs, and platoons may hinder non-CAVs' lane-changing; see Fig.~\ref{fig_background}.
Consequently, platoons may disrupt local traffic flow, which may prevent platooning from attaining its maximal benefits; see \cite{jin2020analysis} for quantification of such compromise.

In this paper, we develop a hybrid queuing model as well as an easy-to-use training algorithm specifically for highways with a mixture of CAV platoons and non-CAV flow.
The model predicts the CAV and non-CAV states based on the traffic demand and highway parameters.
We develop a simulation-based gradient descent algorithm to train the model parameters in an online manner.
We show that the proposed training algorithm can attain a prediction error less than 15\% (resp. 25\%) in a stationary (non-stationary) setting with respect to results given by the Simulation of Urban Mobility (SUMO \cite{behrisch2011sumo}).
We also show that the online trained model can be used for platoon headway regulation.

Our hybrid queuing model (HQM) consists of two traffic classes: CAVs and non-CAVs. CAVs arrive at the system as discrete batches (platoons), and non-CAVs arrive as a continuous flow. CAV platoons are ``condensed'' traffic: a platoon of $n$ CAVs is treated equivalently as $\eta n$ non-CAVs, where $\eta\in(0,1)$ captures the impact of reduced spacing between CAVs. A bottleneck restricts the discharge to downstream. Traffic arrives at the bottleneck after traversing the distance from the highway entrance to the bottleneck, and a constant, nominal traverse time is assumed.
The vehicle discharge rate is upper-bounded by the highway capacity. The HQM can be viewed as a simplification of a class of more sophisticated models for mixed-autonomy traffic, which are designed for more detailed analysis or control purposes \cite{wu2017emergent,calvert2019Evaluation,martinez2020stochastic,piacentini2020macroscopic}. The HQM can be further abstracted and used for higher-level decisions such as routing \cite{lazar2018routing}. This model is a  significant refinement of a preliminary version proposed some of us in \cite{jin2018modeling}.

Our training algorithm determines model parameters by minimizing the square error for the vehicle states on the highway section. The parameters to be determined include the nominal traverse time, the nominal capacity, and the non-CAV's priority ratio. The algorithm improves the parameters heuristically with constant step sizes. The HQM is simulated to compute the square error with respect to input data. The algorithm terminates when the reduction in square error is less than a threshold. Since the fluid queuing model is simple and the space for parameter search is not very large, the computational efficiency is acceptable.

We show via simulation that the proposed HQM and the training algorithm are valid for control design. We consider the regulation of between-platoon headways in response to real-time traffic condition. The HQM can be used to determine the minimal allowable headway between two platoons, which is critical for the headway regulation policy. We show that the policy based on the HQM is very close to the simulated optimal policy.

The rest of this paper is organized as follows. Section~\ref{sec_model} defines the HQM and formulates the model training problem. Section~\ref{sec_algorithm} presents the training algorithms in both stationary and non-stationary settings. Section~\ref{sec_evaluation} validates the model and evaluates the control algorithm in SUMO. Section~\ref{sec_concluding} gives concluding remarks.

%% file: sections/model.tex
\section{Hybrid Queuing Model}
\label{sec_model}

In this section, we introduce the hybrid queuing model (HQM) and formulate the training problem.

\begin{figure}[hbtp]
    \centering
    \includegraphics[width=0.75\linewidth]{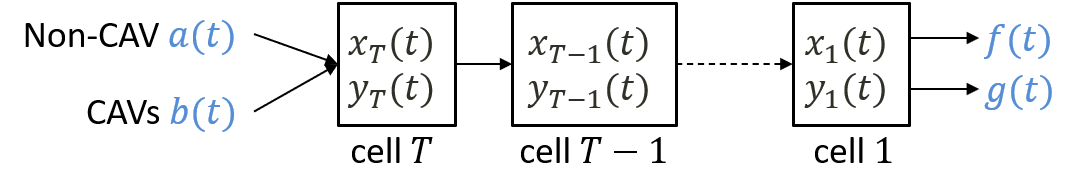}  
    \caption{Hybrid queuing model for a highway section.}
    \label{fig_states}
\end{figure}
We model the generic highway section in Fig.~\ref{fig_background} using the HQM in Fig.~\ref{fig_states}. Cell 1 models the bottleneck.
Cells 2 through $T$ are virtual cells to model the traverse time from the entrance to the bottleneck; the nominal traverse time is $T$.
Every vehicle entering the highway first goes to cell $T$ and then switch from one cell to the immediately downstream one.
No queuing dynamics exists in cells 2 through $T$.
Non-CAVs enter the highway as a continuous traffic flow with rate $a(t)\in\mathbb R_{\ge0}$.
CAVs enter the highway as discrete platoons with size $l$. The number of CAV arrivals during the $t$th time step is $b(t)\in\mathcal B$, where $\mathcal B=\{0,l,2l,\ldots\}$. The mixture of continuous and discrete arrivals leads to the hybrid nature of our model.

The state of the model includes
(i) the non-CAV states $x_k(t)$ in cells $k = 1,2,\ldots,T$ and
(ii) the CAV states $y_k(t)$  in cells $k = 1,2,\ldots,T$.
Hence, the state space is $x(0)\in\mathbb R_{\ge0}^{T}$.
To define the state space, let $\gamma>1$ be a scaling factor for CAVs in a platoon. 
\begin{figure}[hbt]
\centering
\includegraphics[width=0.5\linewidth]{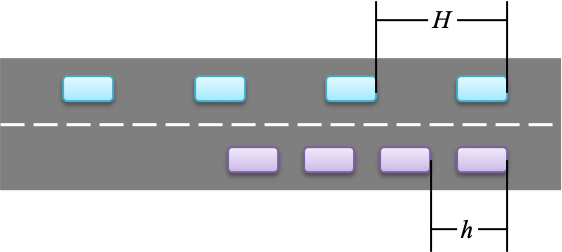}
\caption{$H$ is the spacing between non-CAVs, while $h$ is the spacing within a platoon; $\gamma=H/h$.}
\label{gamma}
\end{figure}
That is, a platoon of $n$ CAVs is equivalent to $n/\gamma$ non-CAVs in terms of its queuing effect; see Fig.~\ref{gamma}.
A typical value for $\gamma$ is 2 \cite{bess+16procieee,santini2016consensus}.
Hence, the set of $x(t)$ is $\mathbb R_{\ge0}^T$, while the set of $y(t)$ is $\mathcal Y^T$, where $\mathcal Y=\{0,l/\gamma,2l/\gamma,\ldots\}$.

Consider an arbitrary initial condition $(x(0),y(0))\in\mathbb R_{\ge0}^{T}\times \mathcal Y^T$.
At the $t$th time step, the state is updated as follows.
\begin{enumerate}
    \item For $k=T$,
    \begin{align*}
        &x_T(t+1)=a(t),\\
        &y_T(t+1)=b(t)/\gamma.
    \end{align*}

    \item For $k=2,3,\ldots,T-1$,
    \begin{align*}
        &x_k(t+1)=x_{k+1}(t),\\
        &y_k(t+1)=y_{k+1}(t).
    \end{align*}
    
    \item For $k=1$, the discharged flows are given by
    \begin{subequations}
    \begin{align}
        &f(t)=\min\Big\{x_1(t),\rho F\Big\},\label{eq_f}\\
        &g(t)=\Big\lfloor\frac 1l\min\Big\{y_1(t),F-f(t)\Big\}\Big\rfloor l,\label{eq_r}
    \end{align}
    \end{subequations}
    where $F\in\mathbb R_{\ge0}$ is the bottleneck's capacity and $\rho\in[0,1]$ specifies the priority of non-CAV traffic with respect to CAVs; $\rho=1$ means that non-CAVs have full priority over CAVs.
    The traffic state in cell 1 is updated as follows.
    \begin{subequations}
        \begin{align*}
       &x_1(t+1)=x_1(t)+x_2(t)-f(t),\\
       &y_1(t+1)=y_1(t)+y_2(t)-g(t).
    \end{align*}
    \end{subequations}
\end{enumerate}

The above queuing dynamics can be interpreted as follows. CAVs and non-CAVs first enter cell $T$ and then enter cell 1 after $(T-1)$ time steps.
No queuing occurs in cells $T,T-1,\ldots,2$, while a two-class traffic queue may exist in cell 1.
The parameters that define the model are the nominal traverse time $T$, the priority $\rho$, and the capacity $F$.
The observable quantities are the inflows $a(t),b(t)$ and the traffic states
$$
\hat m(t)=\|x(t)\|_1,\
\hat n(t)=\|y(t)\|_1,
$$
where $\hat m(t)$ and $\hat n(t)$ stand for number of non-CAV and CAV based on the HQM respectively.

Thus, the estimation problem can be formulated as follows:\\
\indent {\bf Given} $a(s),b(s),m(s),n(s)$, $s=0,1,2,\ldots,t$,\\
\indent {\bf Find} $T,\rho,F$\\
\indent {\bf such that} some cost is minimized.\\
The objective of the HQM is to accurately predict the traffic states $m(t),n(t)$ over the highway section.
Therefore, the cost function penalizes the error between the SUMO-simulated values $\{m(s),n(s);s=0,1,2,\ldots,t\}$ and the predicted values $\{\hat m(s),\hat n(s);s=0,1,2,\ldots,t\}$.
The observed values are obtained from SUMO, while the predicted values are from the HQM.
The specific form of the cost function depends on the specific setting, which we discuss in the next section.

%% file: sections/algorithm.tex
\section{Model training algorithm}
\label{sec_algorithm}

In this section, we consider the model training algorithm in two typical settings, viz. the stationary and the non-stationary settings.
We use 
$$\theta(t)=[T(t),\rho(t),F(t)]$$
to denote the vector of parameters of the HQM. These parameters are the decision variables of the training problem.
Since the state $\hat x(t)$ and $\hat y(t)$ (and thus $\hat m(t)$ and $\hat n(t)$) depends on $\theta$, we write $\hat x(t;\theta)$, $\hat y(t;\theta)$, $\hat m(t;\theta)$, $\hat n(t;\theta)$ to emphasize such dependency.

\begin{figure}[hbt]
\centering
\includegraphics[width=0.48\textwidth]{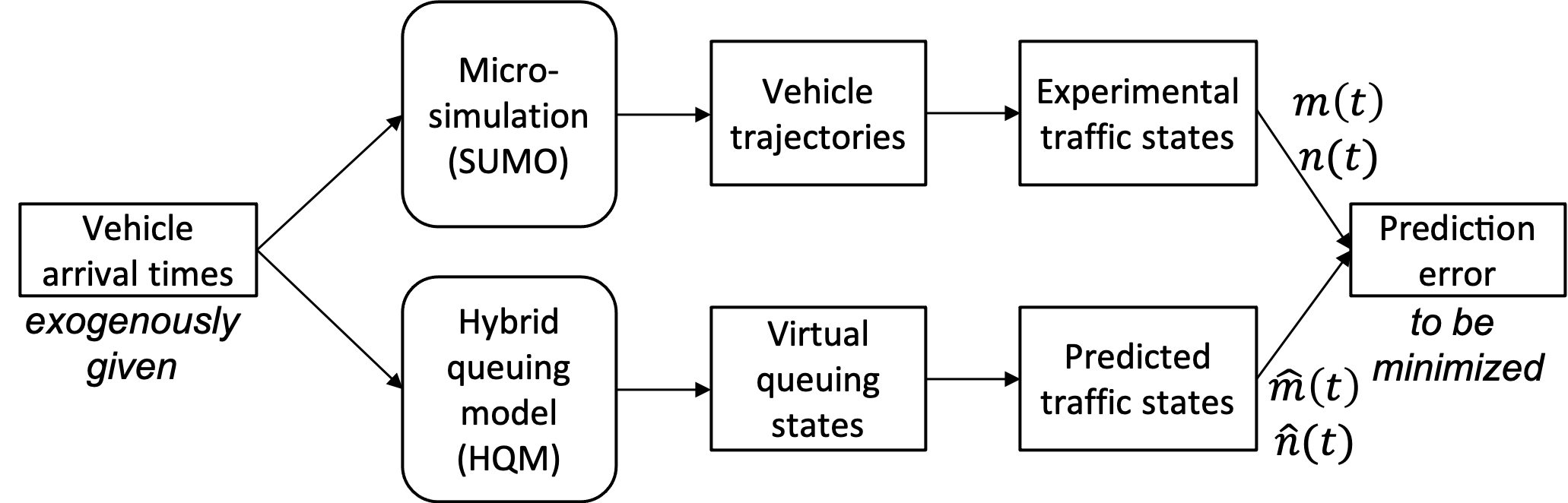}
\caption{HQM approximates SUMO by tracking the vehicle counts on the highway section.}
\label{fig_chart}
\end{figure}

Fig.~\ref{fig_chart} illustrates the mechanism of the training algorithm. SUMO data (vehicle trajectories) are aggregated as CAV and non-CAV counts on the highway section. The HQM uses its internal dynamics to predict the vehicle counts (model prediction), which is supposed to best match the counts (experimental data) given by SUMO. The training algorithm aims to minimizes the prediction error.

In the stationary setting, all parameters are time-invariant, i.e., $\theta(t)=\vartheta$ for all $t\ge0$, where $\vartheta$ is a constant vector. In this setting, SUMO first generates the experimental data first, and then the training algorithm matches the model prediction and the experimental data by varying the model parameters.
In the non-stationary setting, $\theta(t)$ is a non-constant function of $t$. In this setting, as SUMO keeps generating data, the training algorithm computes model prediction and adjusts model parameters simultaneously.
\subsection{Stationary setting}

The objective in this setting is to develop an algorithm that gradually learns the constant parameters and minimizes the prediction error for the traffic count over the highway section.
The one-step cost is given by
\begin{align}
    C(t;\theta)=\Big((\hat m(t;\theta)+\hat n(t;\theta))-(m(t)+n(t))\Big)^2,
    \label{eq_C}
\end{align}
where $m(t)$ and $n(t)$) are the traffic states observed in SUMO.
The cumulative cost is given by
\begin{align*}
    J_1(t;\theta)=\frac1t\sum_{\tau=1}^tC(\tau;\theta).
\end{align*}
The optimal parameter $\theta^*(t)$ is such that
\begin{align*}
    \theta^*(t)=\arg\min_\theta J_1(t;\theta).
\end{align*}
The above optimization problem is solved by a simulation-based heuristic algorithm.
Hence, the estimation $\theta^*(t)$ gradually evolves over time, noticing that $\theta^*$ serves as a parameter for the HQM.

The optimization and iteration are implemented in Algorithm~\ref{alg_optimization}.
Suppose that $\theta^*(0)$ is given according to, say, prior knowledge.
At time $t$, the training algorithm initializes with an initial value $\theta_0(t)=\theta^*(t-1)$.
At the $i$th iteration, we randomly perturb the latest estimate $\theta_{i-1}(t)$ and keep updating until a termination threshold is reached.
\begin{algorithm}[h]
\SetAlgoLined
Initialize $\theta_0(t)=\theta^*(t-1)$\;
\While{$J_1(t;\theta_{i}(t))\le J_1(t;\theta_{i-1}(t))-\epsilon$}{
  Randomly generate a set of perturbations $\mathcal D$ for $\theta$;
  
  $\theta_i(t)=\theta_{i-1}(t)+\arg\min_{\delta\in\mathcal D}J_1(t;\theta_{i-1}(t)+\delta)$;
  }
  \Return $\theta^*(t)=\theta_{i-1}(t)$;
 \caption{Training algorithm in stationary setting}
 \label{alg_optimization}
\end{algorithm}

\subsection{Non-stationary setting}
\label{sub_training2}

The objective in this setting is to develop an algorithm that tracks the drift of the parameters.
The one-step cost $C(s;\theta)$ is again given by \eqref{eq_C}
The cumulative cost is given by
\begin{align*}
    J_2(t;\theta)=\sum_{\tau=1}^t\alpha^{t-\tau}C(\tau;\theta),
\end{align*}
where $\alpha$ stands for the discounted factor.
The optimal parameter $\theta^*(t)$ is such that
\begin{align*}
    \theta^*(t)=\arg\min_\theta J_2(t;\theta).
\end{align*}
The above optimization problem is solved by a simulation-based heuristic algorithm.
Hence, the estimation $\theta^*(t)$ gradually evolves over time.

The optimization and iteration are implemented as follows.
The algorithm starts with some initial value $\theta_0(t)$ according to prior knowledge.
Then, the $i$th estimate $\theta_i(t)$ is randomly perturbed until improved to $\theta_{i+1}(t)$
The termination rule is analogous to that in the stationary case. The pseudo code can be obtained by substituting $J_1$ in Algorithm~\ref{alg_optimization} with $J_2$.

%% file: sections/validation.tex
\section{Evaluation and Validation}
\label{sec_evaluation}

\subsection{Micro-simulation environment}
\begin{figure}[hbtp]
    \centering
    \begin{subfigure}{0.525\linewidth}
    \includegraphics[width=\linewidth]{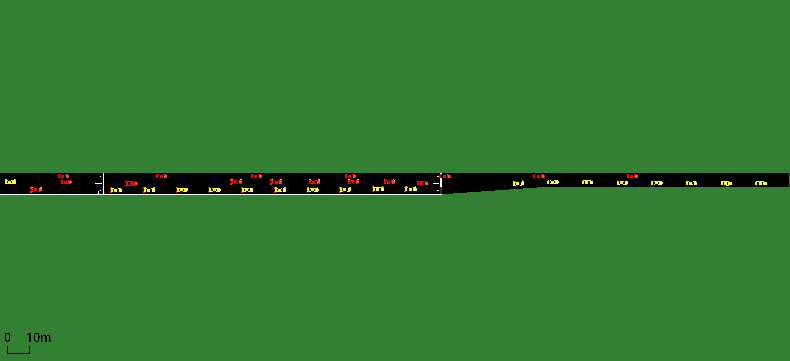}  
    \caption{Part of highway section.}
    \label{highway}
    \end{subfigure}
    \begin{subfigure}{0.41\linewidth}
    \includegraphics[width=\linewidth]{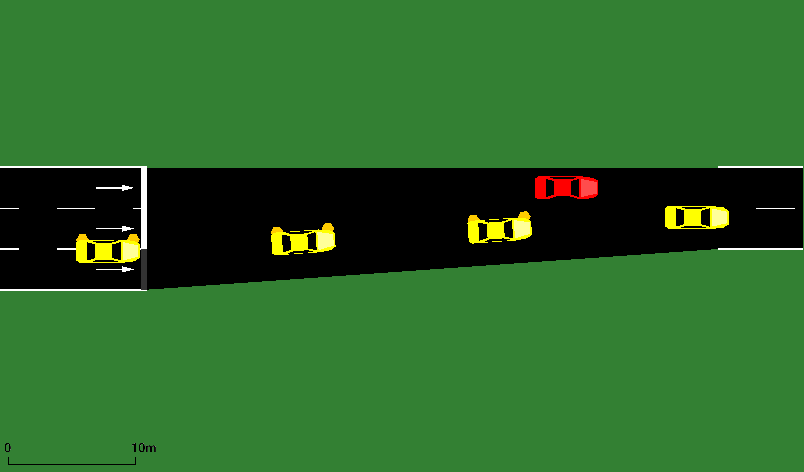}  
    \caption{Bottleneck.}
    \label{configuration}
    \end{subfigure}
    \caption{SUMO simulation testbed. Red (resp. yellow) vehicles are non-CAVs (resp. CAVs).}
    \label{simulation}
\end{figure}

The simulations are conducted in SUMO, see Fig.~\ref{simulation}. SUMO simulations are incorporated with built-in road dynamics, including car-following models and lane-changing models, which realistically represent the ground-truth traffic conditions. For these simulation sets, Gipp's car following model \cite{krauss1998microscopic} and SUMO's lane-changing model \cite{erdmann2014lane} are selected with other SUMO's default settings.

Each individual experimental run simulates 120 minutes of traffic conditions on a highway section. A bottleneck appears where the 3-lane highway section becomes a 2-lane section, delaying the downstream traffic. Based on the traffic demand, vehicle trips are generated randomly to populate the testbed. Generated approaching vehicles are categorized as CAVs and non-CAVs according to CAV penetration rate. To maintain the proposed penetration rate, an arbitrary number of consecutive CAVs are initialized together to populate the highway section. During their trips traversing through this section, they are controlled together as a platoon. In this study, we are adopting the soft platoon state, which means individual CAVs in the same platoon are allowed to appear on separate lanes when the platoon is performing lane changing.

Without applying the proposed control strategies on platoons when they are approaching the highway section, all platoons are set to depart with the nominal travel speed. Taking the advantage of the proposed control scheme, neighboring platoons are designed to distance from each other. When an incoming platoon departs, it is assigned with a guideline velocity based on the difference between the residual time of its leading platoon and itself.

Therefore, under the stationary setting where the traffic demand is fixed during each simulation run, the threshold, based on which a controlled speed applied on the departing platoon, is also a fixed constant. However, for the non-stationary setting under which the traffic demand gradually increase over time until it reaches the desired level, the corresponding threshold also varies to ensure the average travel time of all vehicles is minimized.

\subsection{Stationary setting}

We first apply the training algorithm to the stationary setting. In this setting, traffic enter the highway section with a constant rate.

\begin{table}[hbt]
\caption{Nominal model parameter values.}
\label{tab_parameters}
\centering
\begin{tabular}{@{}lccc@{}}
\toprule
Quantity [unit] & Notation & Nominal value & Learned value \\ 
\hline
Traverse time [sec] & $T$ & 37 & 31 \\ 
Capacity [veh/hr] & $F$ & 3600 & 4000 \\
Scaling factor & $\gamma$ & 2 & 3 \\
Non-CAV priority ratio & $\rho$ & 0.5 & 0.9 \\ 
\bottomrule
\end{tabular}
\end{table}

The results listed in Table~\ref{tab_parameters} are worth discussing. It is interesting that the nominal values are more conservative than the learned values. Specifically, the learned travel time, capacity, and scaling factor are all more optimistic than their respective nominal values. The reason may be that platooning leads to more space on the road, which can further improve the non-CAV traffic flow. In addition, the nominal priority ratio for non-CAVs (0.5) is much smaller than the learned value (0.9); the learned value is very close to one. This implies that without intervention from the RSU, non-CAVs almost enjoy natural priority over CAV platoons. The reason is probably that non-CAVs are more flexible in terms of lane-changing and overtaking maneuvers, while CAV platoons require much more space for those maneuvers.

\begin{figure}[hbt]
    \centering
    \includegraphics[width=0.50\textwidth]{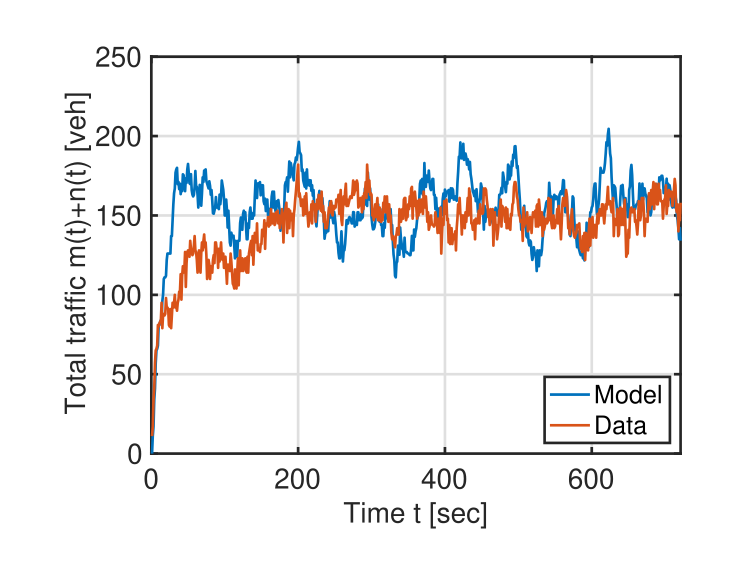}
    \caption{The HQM (``model'') adequately tracks the total vehicle count in SUMO (``data'').}
\label{fig_stationary_compare}
\end{figure}

Fig.~\ref{fig_stationary_compare} shows that the HQM adequately tracks the vehicle count in SUMO. The time-average percentage prediction error is 14.6\%, which is satisfactory given the simplicity of the HQM.
In Fig.~\ref{fig_stationary_compare}, a relatively significant difference is observed during the early stage, where the algorithm starts from initial guess and gradually learns the model parameters. Observed vehicle count based on SUMO simulations demonstrates consistency with the HQM predictions. The oscillations in vehicle count through time by HQM can be illustrated by the lack of explicit road dynamics, such as car-following models and lane-changing models, which usually ``damp'' oscillations. These road dynamics contribute to the relatively stable vehicle counts in the SUMO simulation.

We also use the trained HQM to implement a theoretically optimal headway regulation policy.
This policy ensures that any two consecutive platoons cannot pass the bottleneck within a time interval less than $\Delta$ amount of time. For this policy, $\Delta$ is the major design parameter. The optimal value depending on the HQM is given by
\begin{align*}
    \Delta^*&=\frac{l/\gamma}{F-b}
=\frac{10[veh]/2.5}{4000[veh/hr]-1270[veh/hr]}\\
&=4.4[sec],
\end{align*}
where the value for $b$ is directly obtained from simulation data.
More details about the headway regulation policy are available in \cite{jin2020analysis,cicic2021coordinating}.

\begin{figure}[hbt]
\centering
\includegraphics[width=0.50\textwidth]{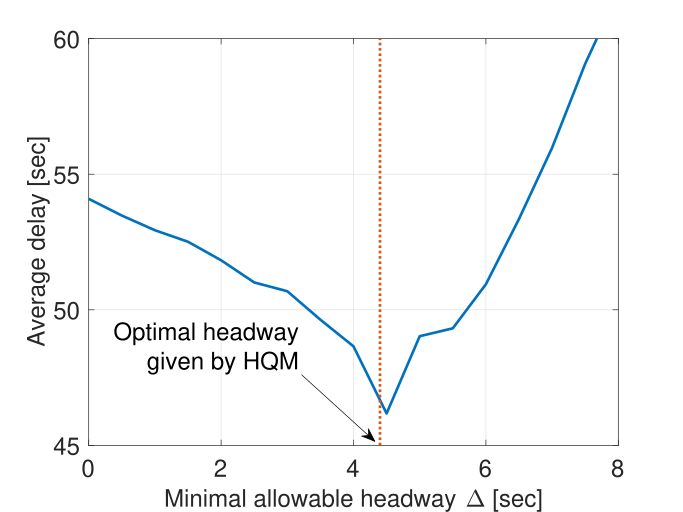}
\caption{Average delay simulated in SUMO. The HQM gives an optimal headway of 4.4 sec, which is close to the numerical optimum.}
\label{fig_TT}
\end{figure}

We apply a range of values for $\Delta$ in SUMO and recorded the average delay; see Fig.~\ref{fig_TT}. The simulated optimal headway is 4.5 sec, which is very close to $\Delta^*$ computed based on the HQM.

\subsection{Non-stationary setting}

In this subsection, we gradually vary the free-flow speed over time and adjust the parameters to track such non-stationarity. Over a 2-hour time interval, the free-flow speed is linearly reduced from 100km/h to 60km/h, which mimics e.g. the change of weather or visibility. The algorithm was described in Section~\ref{sub_training2}



\begin{figure}[hbt]
\centering
\includegraphics[width=0.50\textwidth]{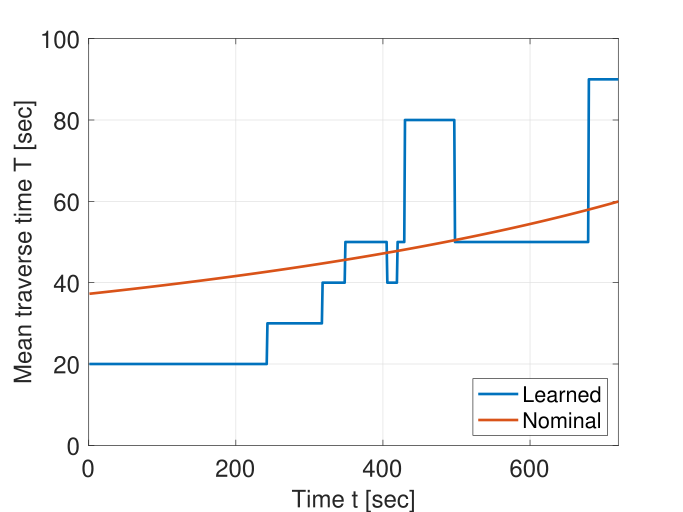}
\caption{The training algorithm adjusts the parameter $T$ to track environmental non-stationarity.}
\label{fig_T}
\end{figure}

Fig.~\ref{fig_T} shows how the learned traverse time $T$ varies over time. The nominal value is given by
$$
T_{\mbox{nominal}}(t)=L/v(t),
$$
where $L$ is the length of the highway section and $v(t)$ is the linearly decreasing free-flow speed.
The learned value tracks the general trend of the nominal value. Similar to the observation in Fig.~\ref{fig_stationary_compare}, the HQM changes more rapidly than SUMO. An explanation is that, since demand is fixed, vehicles may start to interact as free-flow speed decreases.

\begin{figure}[hbt]
\centering
 \includegraphics[width=0.50\textwidth]{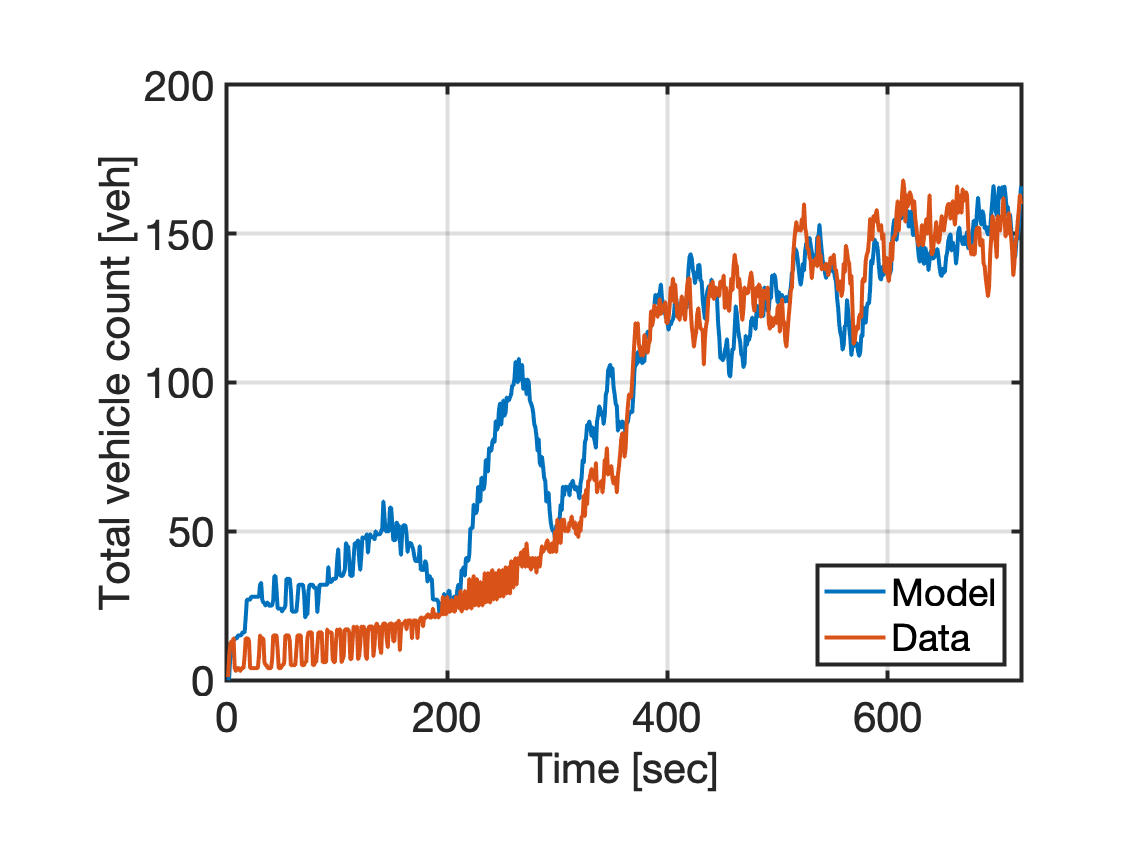}
\caption{HQM gradually learns and tracks SUMO results.}
\label{fig_nonstationary_h12}
\end{figure}

Fig.~\ref{fig_nonstationary_h12} shows that the model provides a good estimation of total vehicle count on the highway. The overall prediction error is 24.38\%, which is higher than the stationary case. However, the error mainly comes from the beginning stage (before 400 sec), where the model parameters are learned from scratch. After 400 sec, the model tracks the vehicle counts much better than at the beginning.

%% file: sections/conclusion.tex
\section{Concluding remarks}
\label{sec_concluding}

In this paper, we considered the training and validation of a hybrid queuing model for highway bottlenecks with CAV platoons. This model describes CAV platoons as discrete bulk arrivals and non-CAV traffic as continuous flow. We studied model training algorithms in both stationary and non-stationary settings. We validate the hybrid queuing model with respect to SUMO results. In the stationary setting, our model gives a prediction error of about 15\%. In the non-stationary setting, the model parameters are able to track the change in the environment, and a 25\% prediction error is attained. The above results provide the basis for analysis and design of learning-based platoon coordination algorithms, which are promising directions for future work.

%% file: Su_cdc.bbl
\begin{thebibliography}{10}
\providecommand{\url}[1]{#1}
\csname url@samestyle\endcsname
\providecommand{\newblock}{\relax}
\providecommand{\bibinfo}[2]{#2}
\providecommand{\BIBentrySTDinterwordspacing}{\spaceskip=0pt\relax}
\providecommand{\BIBentryALTinterwordstretchfactor}{4}
\providecommand{\BIBentryALTinterwordspacing}{\spaceskip=\fontdimen2\font plus
\BIBentryALTinterwordstretchfactor\fontdimen3\font minus
  \fontdimen4\font\relax}
\providecommand{\BIBforeignlanguage}[2]{{%
\expandafter\ifx\csname l@#1\endcsname\relax
\typeout{** WARNING: IEEEtran.bst: No hyphenation pattern has been}%
\typeout{** loaded for the language `#1'. Using the pattern for}%
\typeout{** the default language instead.}%
\else
\language=\csname l@#1\endcsname
\fi
#2}}
\providecommand{\BIBdecl}{\relax}
\BIBdecl

\bibitem{hor+var00}
R.~Horowitz and P.~Varaiya, ``Control design of an automated highway system,''
  \emph{Proceedings of the IEEE}, vol.~88, no.~7, pp. 913--925, 2000.

\bibitem{bess+16procieee}
B.~Besselink, V.~Turri, S.~{van de Hoef}, K.-Y. Liang, A.~Alam,
  J.~M{\aa}rtensson, and K.~H. Johansson, ``Cyber-physical control of road
  freight transport,'' \emph{Proceedings of IEEE}, vol. 104, no.~5, pp.
  1128--1141, 2016.

\bibitem{litman2017autonomous}
T.~Litman, \emph{Autonomous vehicle implementation predictions}.\hskip 1em plus
  0.5em minus 0.4em\relax Victoria Transport Policy Institute Victoria, Canada,
  2017.

\bibitem{naus+10}
G.~J.~L. Naus, R.~P.~A. Vugts, J.~Ploeg, M.~J.~G. van~de Molengraft, and
  M.~Steinbuch, ``String-stable {CACC} design and experimental validation: A
  frequency-domain approach,'' \emph{IEEE Transactions on Vehicular
  Technology}, vol.~59, no.~9, pp. 4268--4279, 2010.

\bibitem{ploeg14}
J.~Ploeg, N.~van~de Wouw, and H.~Nijmeijer, ``{$L_p$} string stability of
  cascaded systems: Application to vehicle platooning,'' \emph{IEEE
  Transactions on Control Systems Technology}, vol.~22, no.~2, pp. 786--793,
  2014.

\bibitem{coogan15interconnected}
S.~Coogan and M.~Arcak, ``A dissipativity approach to safety verification for
  interconnected systems,'' \emph{IEEE Transactions on Automatic Control},
  vol.~60, no.~6, pp. 1722--1727, 2015.

\bibitem{tsugawa16}
S.~Tsugawa, S.~Jeschke, and S.~E. Shladover, ``A review of truck platooning
  projects for energy savings,'' \emph{IEEE Transactions on Intelligent
  Vehicles}, vol.~1, no.~1, pp. 68--77, 2016.

\bibitem{yu2020stability}
H.~Yu, S.~Amin, and M.~Krstic, ``Stability analysis of mixed-autonomy traffic
  with cav platoons using two-class aw-rascle model,'' in \emph{2020 59th IEEE
  Conference on Decision and Control (CDC)}.\hskip 1em plus 0.5em minus
  0.4em\relax IEEE, 2020, pp. 5659--5664.

\bibitem{zhou2020modeling}
J.~Zhou and F.~Zhu, ``Modeling the fundamental diagram of mixed human-driven
  and connected automated vehicles,'' \emph{Transportation research part C:
  emerging technologies}, vol. 115, p. 102614, 2020.

\bibitem{wu2017emergent}
C.~Wu, A.~Kreidieh, E.~Vinitsky, and A.~M. Bayen, ``Emergent behaviors in
  mixed-autonomy traffic,'' in \emph{Conference on Robot Learning}.\hskip 1em
  plus 0.5em minus 0.4em\relax PMLR, 2017, pp. 398--407.

\bibitem{talebpour2016influence}
A.~Talebpour and H.~S. Mahmassani, ``Influence of connected and autonomous
  vehicles on traffic flow stability and throughput,'' \emph{Transportation
  Research Part C: Emerging Technologies}, vol.~71, pp. 143--163, 2016.

\bibitem{jin2020analysis}
L.~Jin, M.~\v{C}i\v{c}i\'{c}, K.~H. Johansson, and S.~Amin, ``Analysis and
  design of vehicle platooning operations on mixed-traffic highways,''
  \emph{IEEE Transactions on Automatic Control}, 2020, available at
  \url{https://ieeexplore.ieee.org/document/9246221}.

\bibitem{behrisch2011sumo}
M.~Behrisch, L.~Bieker, J.~Erdmann, and D.~Krajzewicz, ``Sumo--simulation of
  urban mobility,'' in \emph{The Third International Conference on Advances in
  System Simulation (SIMUL 2011), Barcelona, Spain}, vol.~42, 2011.

\bibitem{calvert2019Evaluation}
S.~C. Calvert, W.~J. Schakel, and B.~V. Arem, ``Evaluation and modelling of the
  traffic flow effects of truck platooning,'' \emph{Transportation Research
  Part C: Emerging Technologies}, vol. 105, no. AUG., pp. 1--22, 2019.

\bibitem{martinez2020stochastic}
I.~Martínez and W.-L. Jin, ``Stochastic lwr model with heterogeneous vehicles:
  Theory and application for autonomous vehicles,'' \emph{Transportation
  Research Procedia}, vol.~47, pp. 155--162, 2020.

\bibitem{piacentini2020macroscopic}
G.~Piacentini, P.~Goatin, and A.~Ferrara, ``A macroscopic model for platooning
  in highway traffic,'' \emph{SIAM Journal on Applied Mathematics}, vol.~80,
  no.~1, pp. 639--656, 2020.

\bibitem{lazar2018routing}
D.~{Lazar}, S.~{Coogan}, and R.~{Pedarsani}, ``Routing for traffic networks
  with mixed autonomy,'' \emph{IEEE Transactions on Automatic Control}, pp.
  1--1, 2020.

\bibitem{jin2018modeling}
L.~Jin, M.~{\v{C}}i{\v{c}}i{\'{c}}, S.~Amin, and K.~H. Johansson, ``Modeling
  the impact of vehicle platooning on highway congestion: A fluid queuing
  approach,'' in \emph{Proceedings of the 21st International Conference on
  Hybrid Systems: Computation and Control (part of CPS Week)}.\hskip 1em plus
  0.5em minus 0.4em\relax ACM, 2018, pp. 237--246.

\bibitem{santini2016consensus}
S.~Santini, A.~Salvi, A.~S. Valente, A.~Pescap{\'e}, M.~Segata, and R.~L.
  Cigno, ``A consensus-based approach for platooning with intervehicular
  communications and its validation in realistic scenarios,'' \emph{IEEE
  Transactions on Vehicular Technology}, vol.~66, no.~3, pp. 1985--1999, 2016.

\bibitem{krauss1998microscopic}
S.~Krau{\ss}, ``Microscopic modeling of traffic flow: Investigation of
  collision free vehicle dynamics,'' 1998.

\bibitem{erdmann2014lane}
J.~Erdmann, ``Lane-changing model in sumo,'' \emph{Proceedings of the SUMO2014
  modeling mobility with open data}, vol.~24, pp. 77--88, 2014.

\bibitem{cicic2021coordinating}
M.~\v{C}i\v{c}i\'c, L.~Jin, and K.~H. Johansson, ``Coordinating vehicle
  platoons for highway bottleneck decongestion and throughput improvement,''
  \emph{IEEE Transactions on Intelligent Transportation Systems}, 2021,
  accepted, available at https://arxiv.org/pdf/1907.13049.pdf.

\end{thebibliography}
